\documentclass[12pt,useAMS,usenatbib,usegraphicx]{mn2e}
\usepackage{wrapfig}
\title[Possible detection of interstellar benzonitrile]{Possible detection of interstellar benzonitrile.}
\author[S. V. Kalenskii]{
  S. V. Kalenskii\thanks{E-mail:kalensky@asc.rssi.ru (SVK)}; 
  \\
  Astro Space Center, Lebedev Physical Institute, 
     Profsoyuznaya 84/32, Moscow, 117997, Russia\\
}

\begin{document}
  \maketitle

\begin{abstract}
The simplest cyanobenzene, benzonitrile (c-C$_6$H$_5$CN) have been possibly detected toward the cyanopolyyne peak in TMC-1. We used the results of the 8.8 -- 50~GHz spectral survey of TMC-1 by Kaifu et al. (2004) and stacked the lines of benzonitrile that fall within the range of this survey. The obtained spectrum strongly suggests the presence of this molecule. Benzonitrile is a derivative of the simplest aromatic hydrocarbon benzene. Aromatic hydrocarbons are thought to be ubiquitous in the ISM, but it is difficult to study them in molecular cloud interiors, since they are nonpolar and have no allowed transitions at radio frequencies. Therefore it is important to search for their derivatives, such as cyanobenzenes. Thus, the detection of benzonitrile might be important for astrochemistry, but additional sensitive observations are necessary in order to confirm it.
\end{abstract}


\section*{Introduction}
It is well known that the most of known cosmic molecules have been found as a result of radio astronomy observations. Among these are complex organic molecules such as glycolaldehyde, 
ethylene glycol, ethyl formate and so on.

While astronomers searched for simple and fairly abundant molecules, such as CO, CS, SiO etc it was sufficient to find only one spectral line with an appropriate frequency in order to make a robust detection of the molecule. Searches for more complex and less abundant molecules, such as methyl formate or glycolaldehyde, which have much weaker lines, required the detection of several lines. With the further increase of sensitivity it was found that, starting from the millimeter waves, the whole frequency range is occupied by weak lines of different molecules. As a result, it is possible to find a line at virtually any frequency. This fact led to spurious detections of glycine and some other molecules. On the other hand, if the abundance of a molecule is no higher than $\sim 10^{-12}$, even the strongest lines of this molecule prove to be buried in this forest of lines, making the detection of such molecule impossible.

At lower frequencies molecular lines are generally much weaker and the lines of molecules with low abundances cannot be detected due to noise in time about several hours even with the modern highly sensitive receivers. Hence, the molecules  that have abundances below than about ten to the minus twelth degree cannot be detected by usual methods both at low and at high frequencies.

\section*{Composite averages}
When individual molecular lines are not seen due to noise or due to the line forest, the molecule can probably be found using so-called {\em composite averages}. This method uses stacking of many lines of the same molecule and is efficient when the observed range of frequencies is broad -- for example, in the case of spectral scans. The method was first described in the paper by~\cite{johan}; a more elaborated version is presented by~\cite{kandja}. 

To build a composite average (CA) one should perform several operations:

\begin{enumerate}
\item
Find the frequencies of the lines of the sought molecule and clip from the initial broadband spectrum narrow-band ({\em elementary}) spectra centered to these frequencies.
\item
Estimate the ratios of the line brightness temperatures and choose the strongest line as the reference one.
\item
Multiply each elementary spectrum by the ratio between the brightness temperatures of the current and the reference line; as a result, the brightness temperatures of all lines will become the same, but the noise root-mean-squares of the elementary spectra (except the reference one) will increase. 
\item
Combine the elementary spectra with weights inversely proportional to their root-mean-squares.
\end{enumerate}

Note that a similar procedure have been successfully used, for example, to search for radio recombination lines (e.g.,~\citet{k84}).

The main problem is to determine the ratios of brightness temperatures. We usually assume that the lines are optically thin and the distribution of energy level populations corresponds to a single temperature. Since this temperature is unknown, we build CAs for a sample of temperatures. However, the latter assumption hardly holds in the case of complex molecules, which can lead to the failure of an attempt to detect such molecule (see below).

\section*{Previous results}

\begin{table}
	\centering
\caption{List of molecules detected by Kalenskii and Johansson~(2010a,2010b) using CAs.}
	\label{tab1}
\begin{tabular}{|l|l|l|}
\hline
Source     & Formula        & Name \\
\hline
DR21(OH)   & CH$_3$OCHO     & Methyl formate  \\
DR21(OH)   & CH$_3$OCH$_3$  & Dimethyl ether  \\
DR21(OH)   & C$_2$H$_5$OH   & Ethanol \\
\hline
W51        & $^{13}$CCH     & Ethynyl; isotopolog\\
           &                &  with $^{13}$C on C1 \\
W51        & CH$_3$NH$_2$   & Methylamine   \\
W51        & (CH$_2$OH)$_2$ & Ethylene glycol \\
W51        & c-C$_2$H$_4$O  & Ethylene oxide  \\
W51        & C$_2$H$_5$OOCH & Ethyl formate   \\
\hline
\end{tabular}
\end{table}

We applied CAs to the analysis of spectral surveys of star-forming regions DR21(OH) and W51e1/e2 performed in the 3-mm wave range (\cite{kandja,kandjb}). In each source we could find several molecules, which were {\em not found} there by traditional methods. They are presented in Table~\ref{tab1}. 
But our main goal was to search for new molecules, and no molecule new for interstellar medium was found as a result of these surveys.

\section*{Benzonitrile}
\begin{figure}
\bigskip
\includegraphics[width=0.5\textwidth]{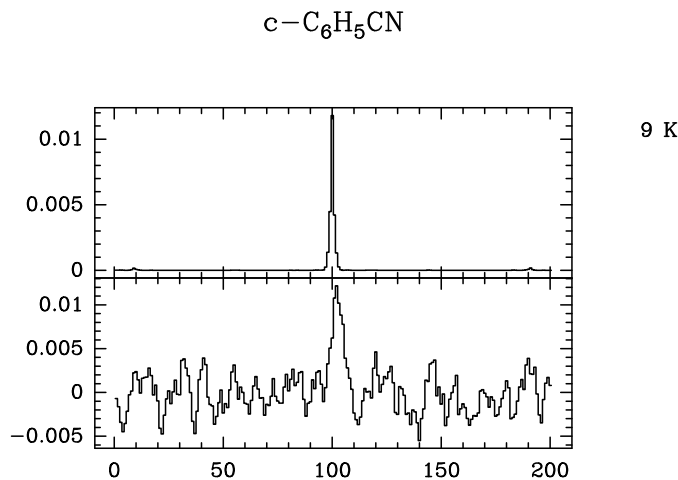}
\caption{Lower panel--CA for benzonitrile (c-C$_6$H$_5$CN) in TMC-1; 
upper panel--expected CA shape in the absence of noise. 
\label{ca1}}
\end{figure}

Wideband spectra of many cosmic objects, obtained in a number of spectral surveys are freely available at the site of project PRIMOS\footnote{http://www.cv.nrao.edu/~aremijan/PRIMOS/} (Prebiotic Interstellar Molecule Survey). In particular, there are the results of a spectral survey of the cyanopolyyne peak in TMC--1, performed with the 45-m Nobeyama radio telescope in the range of frequencies from 8.8 to 50 GHz~(\cite{kaifu}). We built composite averages for a number of molecules, whose spectral data are presented in the  JPL\footnote{http://spec.jpl.nasa.gov} or Cologne\footnote{http://www.astro.uni-koeln.de/cdms/catalog} catalogs of spectral lines and found that the composite average for benzonitrile (C$_6$H$_5$CN) takes the form which suggests that benzonitrile do exist in TMC-1 (Fig.~\ref{ca1}). Here one can clearly see a line just in the center of the spectrum (composite line below).
\begin{figure}
\hspace{2cm}
\includegraphics[width=0.7\textwidth]{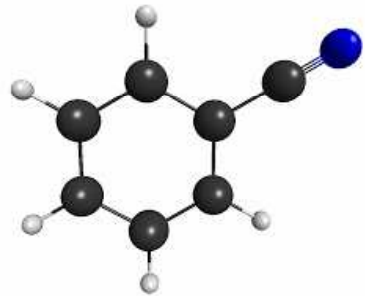}

\vspace{-135mm}
\caption{The molecule of benzonitrile.
\label{bnstr}}
\end{figure}

This detection might open new ways of exploration of aromatic molecules in the ISM. The simplest aromatic hydrocarbon, benzene (C$_6$H$_6$), was detected in the atmosphere of protoplanetary nebula CRL618 more than 15 years ago by~\cite{cern} with the Infrared Space Observatiory (ISO).  Molecules of this type are thought to be ubiguitous in the ISM, but it is difficult to study them in molecular cloud interiors, since they are nonpolar and have no allowed transitions at radio frequencies. Therefore it is important to search for benzene derivatives, such as cyanobenzenes. These compounds are good candidates for the search of aromatic molecules in the ISM due to their large dipole moments, and just benzonitrile is the simplest molecule among them. It is a derivative of benzene, with CN substituted for one of H atoms. 

Thus, the detection of benzonitrile might be important for astrochemistry,  but at this point it is impossible to state that this detection is robust. Nobody have seriously analysed composite averages so far and nobody knows which pitfalls may be encountered. So, how can we check the detection?

\begin{figure}
	\hspace{-1cm}
	\includegraphics[width=0.4\textwidth, angle=-90 ]{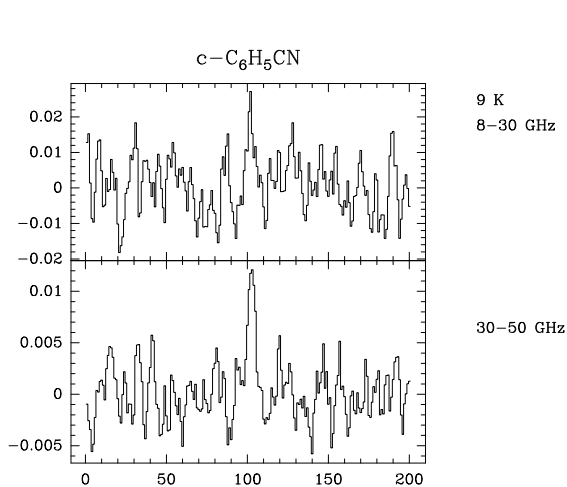}
	\caption{CAs for benzonitrile in different spectral ranges. Upper panel: 8--30~GHz;
lower panel: 30--50 GHz.
\label{spran}}
\end{figure}
\cite{snyder} published essential criteria for establishing the identification of a new interstellar molecule, but these criteria were elaborated for standard methods and are not applicable in the case of composite averages. Therefore we need new specific tests. As the first step we checked each individual spectrum against strong spectral lines or interference spikes that could lead to the appearance of a spurious composite line and found nothing. We made several additional tests and in all cases the behavior of the composite average was just what is expected if the composite line appears as a result of the presence of benzonitrile.

The simplest test was to split the overall frequency range in two halves and build a CA for each half. The result is shown in Fig.~\ref{spran}. The composite lines are seen in both these CAs.

\begin{figure}	
	\centering
\includegraphics[width=0.5\textwidth]{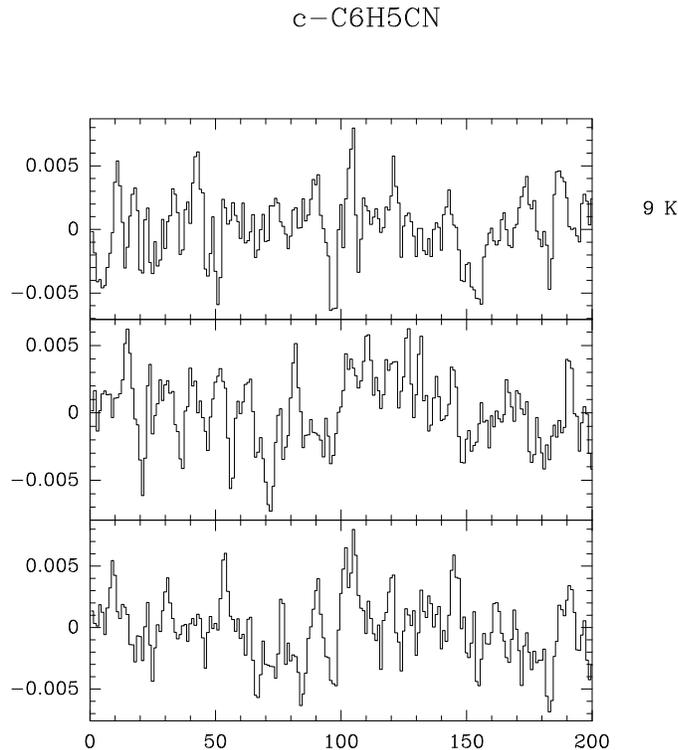} 
\caption{CAs for benzonitrile, built using randomly chosen weights of
the elementary spectra.
\label{rnd}}
\end{figure}

Another test was to build a composite average using randomly chosen weights of the elementary spectra instead of the correct ones. In this case the weights of weak lines increase and those of strong lines decrease, leading to weakening or vanishing of the composite line. However, if the composite line is spurious, its behaviour in this situation is uncertain. We built a number of CAs with randomly chosen weights and the composite line always vanished or nearly vanished. Three so built CAs are shown in Fig.~\ref{rnd}.

This result clearly shows the role of correctly chosen weights of individual spectra. Probably
no new molecule have been found so far using CAs because our assumption that the distribution of energy level populations corresponds to a single temperature rarely holds for complex molecules in the ISM.

The procedure of building composite averages yields the brightness temperature of the composite line equal to that of the strongest molecular line. This means that the brightness temperatures of the strongest lines of benzonitrile are about 0.01 K and they can be observed with the modern receivers. Therefore just sensitive observations of the strongest lines of benzonitrile will be the best test for our results.

\begin{figure}	
	\centering
	\includegraphics[width=0.4\textwidth]{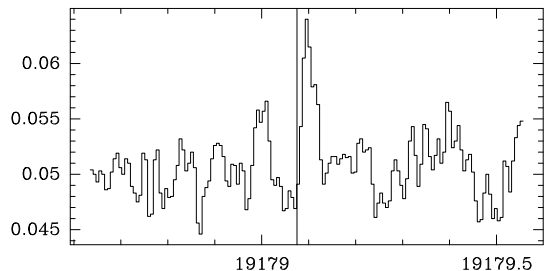}
	\caption{Possible spectral line of benzonitrile.
\label{ind}}
\end{figure}

As the first step in this direction we examined the results of the observations of TMC-1 with the Green Bank telescope, performed several years ago in a number of irregularly spaced frequency ranges 
between 18 and 26~GHz, and found that two relatively strong transitions of benzonitrile fall within the observed ranges. The noise level around the frequency of one of these lines is pretty high, while at the frequency of the other line it is much lower and the line is really visible (Fig.~\ref{ind}). We could not make any identification except benzonitrile for this spectral feature.

Nevertheless, the detection of only one spectral feature with suitable frequency is insufficient to make sure that the molecule is found. Therefore we consider that it is necessary to verify this detection by subsequent sensitive observations of several lines of this molecule. We hope that such observations will confirm the detection.

\section*{Conclusion}
\begin{enumerate}
\item
Composite averages strongly suggest that interstellar benzonitrile exists
in TMC-1. This suggestion should be checked by observing the strongest lines of benzonitrile with a high sensitivity.
\item
In addition to the possible detection of a new important molecule, this result shows that composite averages may be a useful tool to search for new molecules in the ISM.
\end{enumerate}

The work was partly supported by the RFBR grant no.~15-02-07676.

\end{document}